\documentclass[a4paper,12pt]{article}
\usepackage{amssymb,amsmath}
\usepackage{bm}%
\usepackage{stmaryrd}

\def\beq{\begin{equation}}
\def\eeq{\end{equation}}
\def\bea{\begin{eqnarray}}
\def\eea{\end{eqnarray}}
\def\nn{\nonumber}

\def\hf{\frac{1}{2}}

\def\Zgr{$ {\mathbb Z}_2$}
\def\Z2{${\mathbb Z}_2 \times {\mathbb Z}_2$}
\def\g{{\mathfrak g}}

\def\F#1{\frac{\partial}{\partial #1}}
\def\Gl{ {\cal G}_{\ell}}
\def\tGl{ \tilde{\cal G}_{\ell}}
\def\com{\check{\omega}}
\def\BN#1#2{ \left[ \begin{array}{c} #1 \\ #2 \end{array} \right]}
\def\sfP{ \mathsf{\hat{P}}_{ij} }
\def\sfX{ \mathsf{\hat{X}}_{ij} }
\def\sfL{ \mathsf{\hat{\Lambda}}_{ij} }

\begin{document}
%
%
%
%
\thispagestyle{empty}
\vspace*{3cm}
\begin{flushright}
  \today
\end{flushright}
\begin{center}
\textbf{\Large \Z2 generalizations of ${\cal N} = 1$  superconformal Galilei algebras and their representations} \\
\vskip 1cm
N. Aizawa${}^1$, P. S. Isaac${}^2$ and J. Segar${}^3$

\vspace{10mm}

\textit{${}^1$Department of Physical Science, Graduate School of Science,}

\textit{Osaka Prefecture University, Nakamozu Campus,}

\textit{Sakai, Osaka 599-8531 Japan}

\bigskip
\textit{${}^2$ School of Mathematics and Physics,}

\textit{The University of Queensland, St Lucia QLD 4072, Australia}

\bigskip
\textit{${}^3$ Department of Physics, }

\textit{Ramakrishna Mission Vivekananda College, }

\textit{Mylapore, Chennai 600 004, India}
\end{center}

\vfill
\begin{abstract}

  We introduce two classes of novel color superalgebras of \Z2 grading. 
This is done by realizing members of each in the universal enveloping algebra of the ${\cal N}=1$
supersymmetric extension of the conformal Galilei algebra. 
This allows us to upgrade any representation of the super conformal Galilei algebras to a
representation of the \Z2 graded algebra. As an example, boson-fermion Fock space representation
of one class is given. 
We also provide a vector field realization of members of the other class by using a generalization
of the Grassmann calculus to \Z2 graded setting. 

\end{abstract}

\clearpage
\setcounter{page}{1}
%
%
%
\section{Introduction}

Continuous symmetry can be considered as one of the most fundamental concepts in physics and mathematics. 
It is usual to use Lie groups and Lie supergroups to describe symmetries, however, they are not the
only mathematical structure for that purpose. 
Indeed, quantum groups have been used widely in various physical and mathematical problems. 
Before quantum groups, there were attempts to generalize the notion of grading used in the
definition of Lie superalgebras \cite{Ree,rw1,rw2,sch}. 
The basic idea is to generalize the $\mathbb{Z}_2$ graded structure of a Lie superalgebra to  
$ \mathbb{Z}_N, \; \mathbb{Z}_N \times \mathbb{Z}_N,$ or in fact any abelian group. 
The result is a vector space with a specific graded structure and a generalized Lie bracket. 
Nowadays, this generalization of superalgebra is called \textit{color} (super)algebra in the literature. 

The color (super)algebras have been an objects of interest in mathematics for the last four decades,
and many works such as classification of the algebras under some conditions, representations, cohomology and so on, 
have been done \cite{GrJa,sch2,sch3,SchZha,Sil,ChSiVO,SigSil,PionSil,CART,NAJS,NA1,NA2,StoVDJ} (and
references therein). On the other hand, physical applications are very limited
\cite{lr,vas,jyw,zhe,tol2,tol,aktt1,aktt2} so one may conclude that color (super)algebras are
not widely known in the physics community. 

Recently, it was revealed that symmetries of the L\'evy-Leblond equation are given by a color
superalgebra of \Z2 grading \cite{aktt1,aktt2}.  
The L\'evy-Leblond equation is a quantum mechanical wave equation (first order partial differential
equation for a four-component spinor) for a spin 1/2 particle in non-relativistic setting
\cite{LLE}. 
The fact that such a simple and physically important equation has a color supersymmetry suggests
that color superalgebras could be related to more problems in physics. 
In order to clarify which physical systems are related to color superalgebras, 
it is desirable to understand the structure and representations of those color superalgebras which
are, in a sense to be discussed in the current work, connected to certain Lie (super)algebras. 
This is indeed the case of symmetries of the L\'evy-Leblond equation where the color superalgebra has a
connection with super Schr\"odinger algebra. 

Motivated by this, in the present work we consider color superalgebras of \Z2 grading 
which relate to the class of ${\cal N}=1$ superconformal Galilei algebras (SCGA). 
The conformal Galilei algebras (CGA) are infinitesimal generators of conformal symmetry of
non-relativistic space-time \cite{HavPle,NeOlRM}. 
The CGA is defined for arbitrary dimensional space-time and there exist, for a given dimension of
space-time, infinitely many finite-dimensional Lie algebras parameterized by $\ell$
taking a value in non-negative integers or half-integers. 
For example, $\ell = 1/2$ gives the subfamily referred to as the Schr\"odinger algebras. 
Supersymmetric extensions of the CGA have been discussed in many articles
\cite{dh,BecHu,LeLoMin,SY1,SY2,deAzLuk,NaSaYo,Sakaguchi,BagMan,FedLuk,Naru1,Dmod,Mas,Mas2}. 
This is because there does not exist a systematic way of constructing supersymmetric extensions of CGA. 
The present work deals with ${\cal N}=1$ SCGA with arbitrary $\ell$ defined in $(1+1)$ dimensional space-time. 

We present two new results. First, we introduce two classes of novel color superalgebras which are
realized in the universal enveloping algebra of the SCGA. The case of $\ell=1/2$ is already discussed in
\cite{NAJS}. Thus the present result is an extension of \cite{NAJS} to arbitrary $\ell.$ 
Second, we give a vector field representation of the color superalgebra in the space of functions
defined on the color supergroup manifold. This will be helpful when applying the color superalgebra to
physical and mathematical problems and also in considering geometrical properties of the color
supermanifolds themselves.

The plan of this paper is as follows. 
In the next section we recall the definition of a color superalgebra with \Z2 grading, and the
$ {\cal N}=1$ SCGA with and without central extension. From the SCGA we construct two new classes of
color superalgebras in \S \ref{SEC:newZ2Z2}. 
A triangular type decomposition and graded anti-involution of the newly introduced color
superalgebras are also discussed. This allows us to generalize a superstar representation of Lie
superalgebras to the \Z2 graded setting. 
As examples of representations of the color superalgebras, a realization of boson-fermion operators
and a vector field representation on the color supergroup are presented in \S \ref{SEC:Reps}.

%
%
%
\setcounter{equation}{0}
\section{Preliminaries}

\subsection{Color superalgebra of \Z2 grading}

Here we give the definition of color superalgebra of \Z2 grading \cite{rw1,rw2}. 
Let $ \g $ be a vector space and $ \bm{a} = (a_1, a_2)$ an element of  \Z2. 
Suppose that $ \g $ is a direct sum of graded components:
\beq
   \g = \bigoplus_{\bm{a}} \g_{\bm{a}} = \g_{(0,0)} \oplus \g_{(0,1)} \oplus \g_{(1,0)} \oplus \g_{(1,1)}.
\eeq
Homogeneous elements of $ \g_{\bm{a}} $ are denoted by $ X_{\bm{a}}, Y_{\bm{a}}, \dots $ 
If $\g$ admits a bilinear operation (the general Lie bracket), denoted by $ \llbracket \cdot, \cdot \rrbracket, $ 
satisfying the identities:
\bea
  && \llbracket X_{\bm{a}}, Y_{\bm{b}} \rrbracket \in \g_{\bm{a}+\bm{b}}
  \\[3pt]
  && \llbracket X_{\bm{a}}, Y_{\bm{b}} \rrbracket = -(-1)^{\bm{a}\cdot \bm{b}} \llbracket Y_{\bm{b}}, X_{\bm{a}} \rrbracket,
  \\[3pt]
  && (-1)^{\bm{a}\cdot\bm{c}} \llbracket X_{\bm{a}}, \llbracket Y_{\bm{b}}, Z_{\bm{c}} \rrbracket \rrbracket
    + (-1)^{\bm{b}\cdot\bm{a}} \llbracket Y_{\bm{b}}, \llbracket Z_{\bm{c}}, X_{\bm{a}} \rrbracket \rrbracket
    + (-1)^{\bm{c}\cdot\bm{b}} \llbracket Z_{\bm{c}}, \llbracket X_{\bm{a}}, Y_{\bm{b}} \rrbracket \rrbracket =0 
    \label{gradedJ}
\eea
where
\beq
  \bm{a} + \bm{b} = (a_1+b_1, a_2+b_2) \in {\mathbb Z}_2 \times {\mathbb Z}_2, \qquad \bm{a}\cdot \bm{b} = a_1 b_1 + a_2 b_2,
\eeq
then $\g$ is referred to as a color superalgebra of \Z2 grading. 

We take $\g$ to be contained in its enveloping algebra, via the identification
\beq
 \llbracket X_{\bm{a}}, Y_{\bm{b}} \rrbracket =  X_{\bm{a}} Y_{\bm{b}} - (-1)^{\bm{a}\cdot \bm{b}}
Y_{\bm{b}} X_{\bm{a}},
\eeq
where an expression such as $ X_{\bm{a}} Y_{\bm{b}}$ is understood to denote the associative product
on the enveloping algebra. 
In other words, by definition, in the enveloping algebra the general Lie bracket $ \llbracket \cdot, \cdot
\rrbracket $ for homogeneous elements coincides with either a commutator or anticommutator. 

 This is a natural generalization of Lie superalgebra which is defined on \Zgr-grading structure:
\beq
  \g = \g_{(0)} \oplus \g_{(1)}
\eeq
with 
\beq
  \bm{a} + \bm{b} = (a+b), \qquad \bm{a} \cdot \bm{b} = ab.
\eeq
It should be noted that $ \g_{(0,0)} \oplus \g_{(0,1)} $ and $ \g_{(0,0)} \oplus \g_{(1,0)} $ are sub-superalgebras of the \Z2-graded 
superalgebra $\g.$

%
\subsection{${\cal N}=1$ superconformal Galilei algebras}

The ${\cal N}=1$ superconformal Galilean algebras is a family of Lie superalgebras  parametrized by $\ell$ which take a value in $ \mathbb{N} $ or $ \mathbb{N} + \hf $ where 
$ \mathbb{N} $ denotes the set of non-negative integers. 
The bosonic elements are denoted by $\{H,D,K,P_n,\}_{n=0,1,\dots,2\ell}$ and the fermionic 
elements by  $\{Q,S,X_n\}_{n=0,1,\dots,2\ell-1}$. 
We present below nonvanishing (anti)commutators of 
the algebra:
\begin{equation}
   \begin{array}{lll}
      [D, H] = H, & [D, K] = -K, & [H, K] = 2D, \\[3pt]
      [D, Q] = \hf Q, &  [K, Q] = S, & [H, S] = Q,                                   
      \\[3pt]
      [D, S] = -\hf S, & \{ Q, Q \} = 2 H, & 
      \{ S, S \} = -2  K, \\[3pt]
      \{ Q, S \} = - 2 D,  &
            [H, P_n] = nP_{n-1}, & [D, P_n] = -(˘n-\ell)P_n, \\[3pt]
             [H, Q] = 0, &
       [K, P_n] = -(n-2\ell)P_{n+1}, & [Q, P_n] =n X_{n-1}, \\[3pt]
        [H , X_n]=nX_{n-1}, & \{Q, X_n\}=P_n, &
        [K, X_n] =-(n-2\ell+1)X_{n+1}, \\[3pt]
         [S, P_n] = (n-2\ell)X_n, & [D, X_n]=-(n-\ell 
         +\hf)X_n, & \  \
         \{S, X_n\} = P_{n+1}.
   \end{array}
   \label{sl21plus}
\end{equation}
We denote this superalgebra by $\g_{\ell}$. 
As is seen from \eqref{sl21plus} $\g_{\ell}$ consists of $ sl(2) $ spanned by $ H, D, K $ and its
modules. $ P_n$ is a spin $ \ell$ module under the adjoint action of $ sl(2),$
$X_n$ is a spin $ \ell-\hf$ module and $ Q, S $ is a spin $\hf$ module.  

The superalgebra $\g_{\ell}$ admits the central extension if $ \ell \in \mathbb{N} + \hf$:
\begin{alignat}{2}
  [P_n, P_m] &= \delta_{m+n,2\ell}\, c\,I_n, & \quad  \{ X_n, X_m  \} &=\delta_{m+n,2\ell-1}\, c\, \alpha_n, 
  \nn \\
  I_n &= (-1)^n n! (2\ell-n)!, & \alpha_n &= I_n/(2\ell-n),
  \label{MassExt}
\end{alignat}
where $c$ is the central charge. 
We denote the algebra with the central extension by $\tilde{\g}_{\ell}.$

%
%
%
\setcounter{equation}{0}
\section{\Z2 generalization of $ \g_{\ell} $ and $\tilde{\g}_{\ell}$}
\label{SEC:newZ2Z2}

\subsection{Defining relations of new color superalgebras}

 We would like to create a color superalgebra of \Z2 grading from the superalgebras $ \g_{\ell} $ and 
$ \tilde{\g}_{\ell}.$ 
Our strategy is to realize the color superalgebra in the universal enveloping algebra of $\g_{\ell}$
or $ \tilde{\g}_{\ell}.$ 
For elements with a fixed grading (i.e. homogeneous elements), the general bracket 
$ \llbracket \ , \ \rrbracket $ is
either symmetric or antisymmetric. Thus we use the standard notation [ , ] for the antisymmetric
bracket and $\{$ , $\}$ for the symmetric bracket.

Let us start with $U(\g_{\ell}),$ the universal enveloping algebra of $\g_{\ell}.$ 
We take the following particular elements of $U(\g_{\ell})$
\begin{equation}
   P _{nm} \equiv \{ P_n, P_m \}, \quad \Lambda_{nm}\equiv\{P_n,X_m\}, \quad 
 X_{nm} \equiv [X_n,X_m] 
 \label{PPdef}
\end{equation}
and consider the vector space spanned by these elements and the basis  of $\g_{\ell}.$  
We assign the \Z2 degree as follows:
\beq
  \begin{array}{ccl}
  	(0,0) & : & H, \ D, \ K, \ P_{nm}, \ X_{nm} \\[3pt]
  	(0,1) & : & P_n \\[3pt]
  	(1,0) & : & Q, \ S,  \ \Lambda_{nm} \\[3pt]
  	(1,1) & : & X_n
  \end{array}
  \label{Z2-121}
\eeq
It is then straightforward to verify, with the aid of \eqref{sl21plus}, that the elements in (\ref{Z2-121}) close in the \Z2 graded 
(anti)commutation relations. 
We give the non-vanishing relations below. 
For the sake of simplicity we set $ c = 1$ in the computation. 

%
%

\medskip\noindent
$(0,0)$-$(0,0)$ sector:
\beq
\arraycolsep=10pt
  \begin{array}{lll}
    [D, H] = H, & [H, K]=2D, \qquad [D, K]=-K, &  \\[3pt]
    \multicolumn{2}{l}{ [H, P_{nm}]=n P_{n-1\; m} + m P_{n\;
     m-1},} \\[3pt]
    \multicolumn{2}{l}{ [H, X_{n m}] = nX_{n-1\; m} + m X_{n\; m-1},} \\[3pt]
     \multicolumn{2}{l}{[D, P_{nm}]=-(n+m-2 \ell )P_{nm},} \\[3pt]
     \multicolumn{2}{l}{[D, X_{n m} ]=-(n+m-2\ell+1)X_{nm},} \\[3pt]
     \multicolumn{2}{l}{[K, P_{nm}]=-(n-2\ell)P_{n+1\; m} -(m-2\ell ) P_{n\; m+1}, } \\[3pt]
     \multicolumn{2}{l}{[K, X_{nm}]=-(n-2\ell +1)X_{n+1\; m} -(m-2\ell +1)X_{n\; m+1}.} 
  \end{array}
   \label{00-00}
\eeq

\noindent
$(0,0)$-$(0,1)$ sector:
\begin{equation}
[H, P_n]=nP_{n-1}, \qquad [D,P_n]=-(n - \ell)P_n , \qquad [K,P_n]=-(n-2\ell) P_{n+1}.
\end{equation}

\noindent
$(0,0)$-$(1,0)$ sector:
\beq
\arraycolsep=10pt
  \begin{array}{lll}
     [H, S] = Q, & [D, Q] = \hf Q, & [D, S] = -\hf S,  \\[3pt]
     [K, Q] =  S, & [H, \Lambda_{nm}]=n\Lambda_{n-1\;m}+m\Lambda_{n\; m-1} , & [D, \Lambda_{nm}] = -(n+m-2\ell +\hf) \Lambda_{nm},  \\[3pt]
     \multicolumn{2}{l}{[K, \Lambda_{nm}]=-(n-2\ell) \Lambda_{n+1 m} -(m-2\ell +1)\Lambda_{nm+1},} \\[3pt]
     \multicolumn{2}{l}{ [Q, P_{nm}]=n\Lambda_{m \;  n-1} +m\Lambda_{n \;  m-1},} \\[3pt]
      \multicolumn{2}{l}{ [Q, X_{nm}]=\Lambda_{n\,  m} -\Lambda_{m \, n},} \\[3pt]
  \multicolumn{2}{l}{ [S, P_{nm}]=(m-2\ell)\Lambda_{n \,  m}  +(n-2\ell){\Lambda_{m\,n}},}\\[3pt]
      \multicolumn{2}{l}{ [S, X_{nm}]=\Lambda_{n+1\;  m} -\Lambda_{m+1 \; n}.} \\[3pt]
   
  \end{array}
\eeq

\medskip\noindent
$(0,0)$-$(1,1)$ sector:
\beq
 [H , X_n]=n X_{n-1},\qquad [D,X_n]=-(n-\ell+\hf) X_n, \qquad [K,X_n]=-(n-2 \ell +1)X_{n+1}.
\eeq

\noindent
$(0,1)$-$(0,1)$ sector:
\beq
  \{P_n, P_m \} = P_{nm}. \
\eeq

\noindent
$(0,1)$-$(1,0)$ sector:
\beq
  [P_n, Q ] = -nX_{n-1}, \qquad [P_n, S] = -(n-2\ell)X_n. 
\eeq

\noindent
$(0,1)$-$(1,1)$ sector:
\beq
  \{ P_n, X_m \} = \Lambda_{nm}.
\eeq

\noindent
$(1,0)$-$(1,0)$ sector:
\begin{equation}
\{Q,Q\}=2H, \qquad \{Q, S\}=-2D, \qquad \{S , S \}=-2K,
\end{equation}
\bea
 \{Q, \Lambda_{nm}\}&=&n X_{n-1\;  m} +P_{n \, m}, \nn \\ \\
\{S, \Lambda_{nm} \}&=&(n-2\ell) X_{n\,  m} +P_{n \; m+1}.  
\eea 
$(1,0)$-$(1,1)$ sector:
\beq
  \{ Q , X_n \} = P_n, \qquad \{ S, X_n \} = P_{n+1}.
\eeq

\noindent
$(1,1)$-$(1,1)$ sector:
\beq
   [X_n , X_m] = X_{nm}.  \label{11-11}
\eeq
In order to assert that these relations define a color superalgebra, 
one must show that the relations are compatible with the graded Jacobi identity without using \eqref{sl21plus}. 
Indeed one may verify the compatibility by the direct but lengthy computation. 
Thus we have obtained a color superalgebra and it is denoted by ${\cal G}_{\ell}.$ 

 Next let us consider the case of $ U(\tilde{\g}_{\ell})$ with $ \ell \in \mathbb{N} + \hf. $  
We show that the vector space spanned by the same elements \eqref{Z2-121} as $ U(\g_{\ell})$ gives another example of color superalgebra. In this case we have more non-vanishing relations in addition to \eqref{00-00}-\eqref{11-11}. 
Those new relations are given below:

\medskip
\noindent
$(0,0)$-$(0,0)$ sector:
\begin{align}
   [P_{nm}, P_{kr}] &= 2 I_n (\delta_{n+k,2\ell} P_{mr} + \delta_{n+r,2\ell} P_{mk})
   + 2 I_m (\delta_{m+k,2\ell} P_{nr} + \delta_{m+r,2\ell} P_{nk}),
   \nn \\[3pt]
   [ X_{nm}, X_{kr} ] &=  2 \alpha_n (-\delta_{n+k,2\ell-1} X_{mr} + \delta_{n+r,2\ell-1} X_{mk})
   \nn \\[3pt]
   &+ 2 \alpha_m (\delta_{m+k,2\ell-1} X_{nr} - \delta_{m+r,2\ell-1} X_{nk}).
\end{align}

\medskip
\noindent
$(0,0)$-$(0,1)$ sector:
\beq
   [P_{nm}, P_k] = 2 \delta_{n+k,2\ell} I_n P_m + 2 \delta_{m+k,2\ell} I_mP_n.
\eeq

\medskip
\noindent
$(0,0)$-$(1,0)$ sector:
\begin{align}
  [P_{nm}, \Lambda_{kr} ] &=  2 \delta_{n+k,2\ell} I_n\Lambda_{mr} + 2 \delta_{m+k,2\ell} I_m \Lambda_{nr},
  \nn \\[3pt]
  [X_{nm}, \Lambda_{kr}] &= -2 \delta_{n+r,2\ell-1} \alpha_n \Lambda_{km} + 2 \delta_{m+r,2\ell-1} \alpha_m \Lambda_{kn}.
\end{align}

\medskip
\noindent
$(0,0)$-$(1,1)$ sector:
\beq
  [X_{nm}, X_k] = -2\delta_{n+k,2\ell-1} \alpha_n X_m + 2 \delta_{m+k,2\ell-1} \alpha_m X_n.
\eeq

\medskip
\noindent
$(0,1)$-$(1,0)$ sector:
\beq
  [P_n, \Lambda_{mk}] = 2 \delta_{n+m,2\ell} I_n X_k.
\eeq

\medskip
\noindent
$(1,0)$-$(1,0)$ sector:
\beq
  \{ \Lambda_{nm}, \Lambda_{kr} \} = 2 \delta_{n+k,2\ell} I_n X_{mr} + 2 \delta_{m+r, 2\ell-1} \alpha_m P_{nk}.
\eeq

\medskip
\noindent
$(1,0)$-$(1,1)$ sector:
\beq
   \{ \Lambda_{nm}, X_k \} =  2 \delta_{m+k,2\ell-1} \alpha_m P_n.
\eeq

 By the direct computation, one may verify the \Z2 graded Jacobi identities without using \eqref{sl21plus} and \eqref{MassExt}. 
We thus have obtained a novel color superalgebra of \Z2 grading and we call it 
$ \tilde{{\cal G}}_{\ell}.$

 Due to the symmetries $ P_{nm} = P_{mn}, X_{nm} = -X_{mn} $ the number of independent elements of $ P_{nm}, X_{nm}, \Lambda_{nm} $ are $ (\ell+1)(2\ell+1), \ell(2\ell-1), 2\ell (2\ell+1),$ respectively. It follows that 
 $ \dim \Gl = \dim \tGl = 8\ell (\ell+1) + 7. $ 

%
\subsection{Some properties of $ {\cal G}_{\ell}$ and $ \tilde{{\cal G}}_{\ell}$}

The color superalgebras $ \Gl$ and $ \tGl$ have a vector space decomposition corresponding to the triangular decomposition of Lie and Lie superalgebras. 
The decomposition  for $ \Gl$ and $ \tGl$ has no difference so that we discuss on $\Gl. $  
The color superalgebra $ \Gl$ is decomposed into three vector subspaces according to the 
eigenvalues of ad$D$:
\begin{equation}
 \Gl = \Gl^+ \oplus \Gl^0 \oplus \Gl^-,
\end{equation}
where
\begin{equation}
    \Gl^{\pm} = \{ Y \in \Gl \ | \ [D, Y] = \lambda Y, \ \lambda \gtrless 0 \}
\end{equation}
and $ \Gl^0 $ is defined as the subspace of zero eigenvalue. 

If $ \ell \in \mathbb{N} $ then $ \Gl^0 $ is spanned by 
\begin{equation}
  D, \quad P_{n\; 2\ell-n} \ \; (0 \leq n \leq \ell), \quad
  X_{n\; 2\ell-1-n} \ \; (0 \leq n \leq \ell-1), \quad P_{\ell},
\end{equation}
while if $ \ell \in \mathbb{N}+\hf $ it is spanned by
\begin{equation}
   D, \quad P_{n\; 2\ell-n} \ \; (0 \leq n \leq \ell-\hf), \quad
  X_{n\; 2\ell-1-n} \ \; (0 \leq n \leq \ell-\frac{3}{2}), \quad  X_{\ell-\hf}.
\end{equation}
The entries of $ \Gl^{\pm}$ are given by
\begin{align}
  \Gl^+ &= \{ \; H,\; P_{nm}, \; X_{nm}, \; P_n, \; Q, \; \Lambda_{nm}, \; X_n \; \},
  \nn
  \\[3pt]
  \Gl^- &= \{ \; K,\; P_{nm}, \; X_{nm}, \; P_n, \; S, \; \Lambda_{nm}, \; X_n \; \},
\end{align}
and the range of indices is summarized as follows. For $ \Gl^+$
\[
  \begin{array}{ccc}
          & \ell \in \mathbb{N} & \ell \in \mathbb{N}+\hf \\ \hline
     P_{nm} \quad & 0 \leq n \leq \ell-1 \qquad &   0 \leq n \leq \ell-\hf  \\[3pt]
             & \multicolumn{2}{c}{n \leq m \leq 2\ell-1-n}
             \\[6pt]
     X_{nm}  &  0 \leq n \leq \ell-2 & 0 \leq n \leq \ell - \frac{3}{2}  \\[3pt]
            & \multicolumn{2}{c}{n+1 \leq m \leq 2\ell-2-n}
            \\[6pt]
     P_n    & 0 \leq n \leq \ell-1 & 0 \leq n \leq \ell-\hf 
            \\[6pt]
     \Lambda_{nm} & \multicolumn{2}{c}{0 \leq n \leq 2\ell-1} \\[3pt] 
                  & \multicolumn{2}{c}{0 \leq m \leq 2\ell-1-n}
     \\[6pt]
     X_n    & 0 \leq n \leq \ell-1 & 0 \leq n \leq \ell-\frac{3}{2}
  \end{array}
\]
and for $ \Gl^-$
\[
  \begin{array}{ccc}
          & \ell \in \mathbb{N} & \ell \in \mathbb{N}+\hf \\ \hline
     P_{nm} \quad & \ell+1 \leq m \leq 2\ell \qquad &   \ell+\hf \leq m \leq 2\ell  \\[3pt]
             & \multicolumn{2}{c}{2\ell+1-m \leq n \leq m}
             \\[6pt]
     X_{nm}  &  \ell+1 \leq m \leq 2\ell-1 & \ell+\hf \leq m \leq 2\ell-1  \\[3pt]
            & \multicolumn{2}{c}{2\ell-m \leq n \leq m-1}
            \\[6pt]
     P_n    & \ell+1 \leq n \leq 2\ell & \ell+\hf \leq n \leq 2\ell 
            \\[6pt]
     \Lambda_{nm} & \multicolumn{2}{c}{1 \leq n \leq 2\ell} \\[3pt] 
                  & \multicolumn{2}{c}{2\ell-n \leq m \leq 2\ell-1}
     \\[6pt]
     X_n    & \ell \leq n \leq 2 \ell-1 & \ell+\hf \leq n \leq 2\ell-1
  \end{array}
\]
It is straightforward to verify that $ [ \Gl^0, \Gl^{\pm} ] \subseteq \Gl^{\pm}. $

  Next we consider (graded) anti-involutions on the color superalgebras that is a natural generalization of the superalgebras. 
We start with a general definition and follow the terminology of \cite{FrScSo}. 
Let $\g$ be a color superalgebra with \Z2 grading and $ X_{\bm{a}}, Y_{\bm{b}} \in \g $ 
with $ \bm{a}, \bm{b} \in $ \Z2. 
We define two (graded) anti-involution on $\g.$   
The first one $\omega : \g \to \g $, called the \textit{adjoint operation} on $\g$, 
is defined by the properties
\begin{enumerate}
   \renewcommand{\labelenumi}{(\roman{enumi})}
   \item $ \omega(X_{\bm{a}}) \in \g_{\bm{a}} $
   \item $ \omega( \alpha X_{\bm{a}} + \beta Y_{\bm{b}}) = \alpha^* \omega(X_{\bm{a}}) + \beta^* \omega(Y_{\bm{b}}), \quad \alpha, \beta \in \mathbb{C}  $
   \item $ \omega(\, \llbracket X_{\bm{a}}, Y_{\bm{b}} \rrbracket\, ) = \llbracket \omega(Y_{\bm{b}}), \omega(X_{\bm{a}}) \rrbracket $
   \item $ \omega (\omega ( X_{\bm{a}})) = X_{\bm{a}} $
\end{enumerate}
where $ \alpha^* $ is the complex conjugation of $ \alpha. $ 

 The second one is the \textit{superadjoint operation} $ \com : \g \to \g $ defined by
\begin{enumerate}
   \renewcommand{\labelenumi}{(\roman{enumi})}
   \item $ \com(X_{\bm{a}}) \in \g_{\bm{a}} $
   \item $ \com( \alpha X_{\bm{a}} + \beta Y_{\bm{b}}) = \alpha^* \com(X_{\bm{a}}) + \beta^* \com(Y_{\bm{b}}), \quad \alpha, \beta \in \mathbb{C}  $
   \item $ \com(\, \llbracket X_{\bm{a}}, Y_{\bm{b}} \rrbracket\, ) = (-1)^{\bm{a}\cdot\bm{b}} \,\llbracket \com(Y_{\bm{b}}), \com(X_{\bm{a}}) \rrbracket $
   \item $ \com (\com ( X_{\bm{a}})) = (-1)^{a_1+a_2}X_{\bm{a}} $
\end{enumerate}
where $ \bm{a} = (a_1,a_2).$ 

These anti-involutions allow us to generalize the star and superstar representations of Lie superalgebras \cite{FrScSo} to color superalgebras. 
Let $\pi$ be a representation of $\g $ acting in a \Z2-graded vector space. 
Then $\pi$ is a \textit{star} representation of $\g$ if $ \pi \circ \omega = \omega \circ \pi, $ 
and a \textit{superstar} representation of $\g$ if $ \pi \circ \com = \com \circ \pi.$

The color superalgebras $\Gl $ and $ \tGl$ admit two adjoint operations for all possible values of $ \ell:$
\begin{align}
   \omega(H) &= -K, & \omega(K) &= -H, & \omega(D) &= D, \nn \\
   \omega(P_{nm}) &= P_{2\ell-n\; 2\ell-m}, & 
   \omega(X_{nm}) &= - X_{2\ell-1-n\; 2\ell-1-m}, & 
   \omega(P_n) &= \pm P_{2\ell-n},
   \nn \\
   \omega(Q)  &= S,
   &
   \omega(S) &= Q, & \omega(\Lambda_{nm}) &= \Lambda_{2\ell-n\;2\ell-1-m},
   \nn \\
   \omega(X_n) &= \pm X_{2\ell-1-n}, \label{adjoint1}
\end{align}
and
\begin{align}
   \omega(H) &= -K, & \omega(K) &= -H, & \omega(D) &= D, \nn \\
   \omega(P_{nm}) &= P_{2\ell-n\; 2\ell-m}, & 
   \omega(X_{nm}) &= - X_{2\ell-1-n\; 2\ell-1-m}, & 
   \omega(P_n) &= \pm P_{2\ell-n},
   \nn \\
   \omega(Q)  &= -S,
   &
   \omega(S) &= -Q, & \omega(\Lambda_{nm}) &= -\Lambda_{2\ell-n\;2\ell-1-m},
   \nn \\
   \omega(X_n) &= \mp X_{2\ell-1-n}.   \label{adjoint2}
\end{align}
On the other hand $ \Gl $ and $ \tGl $ admit the following superadjoint operation for 
$ \ell \in \mathbb{N}+\hf$
\begin{align}
  \com(H) &= K, & \com(K) &= H, \nn \\
  \com(D) &= D,
   &
  \com(P_{nm}) &= (-1)^{n+m+1} P_{2\ell-n\; 2\ell-m}, \nn 
  \\
  \com(X_{nm}) &= (-1)^{n+m+1} X_{2\ell-1-n\; 2\ell-1-m}, 
  &
  \com(P_n) &= \pm (-1)^n P_{2\ell-n},
  \nn \\
  \com(Q) &= S, & \com(S) &= -Q,
  \nn \\
  \com(\Lambda_{nm}) &= (-1)^{n+m} \Lambda_{2\ell-n\;2\ell-1-m}, &
  \com(X_n) &= \mp (-1)^{n} X_{2\ell-1-n}. 
  \label{superadjoint1}
\end{align}

%
%
%
\setcounter{equation}{0}
\section{Representations of $ \Gl$ and $ \tGl$}
\label{SEC:Reps}

In this section we give two examples of representations of $ \Gl $ and $\tGl.$ 

\subsection{Boson-fermion realization of $\tGl$}

As seen in \S \ref{SEC:newZ2Z2} the algebras $\Gl$ and $\tGl$ are realized in the enveloping algebra of SCGA. 
Thus any representation of the SCGA is converted to the corresponding representation of $ \Gl $ or $ \tGl. $  
As an example, we give a realization of SCGA  with the mass central extension $ \tilde{\g}_{\ell} $ in terms of boson and fermion operators. 
This allows us to define a Fock representation of SCGA  and $\tGl. $ 
To this end, we note that \eqref{MassExt} is nothing but the boson and fermion operators. 
Therefore, it is enough to realize $ \tilde{\g}_{\ell} $ in terms of $ P_n $ and $ X_n $ 
(see \cite{Dmod,Naru1} for ${\cal N}=2 $ SCGA).   

\begin{align}
  H &= -\hf \left(
    \sum_{n=1}^{2\ell} \frac{n}{I_n} P_{2\ell-n} P_{n-1} + \sum_{n=1}^{2\ell-1} \frac{n}{\alpha_n} X_{2\ell-1-n} X_{n-1}
  \right),
  \nn \\
  D &= \hf \left(
    \sum_{n=0}^{2\ell} \frac{n-\ell}{I_n} P_{2\ell-n} P_n 
    + \sum_{n=0}^{2\ell-1} \frac{n+\hf-\ell}{\alpha_n} X_{2\ell-1-n} X_n
  \right),
  \nn \\
  K &= \hf \left(
     \sum_{n=1}^{2\ell} \frac{n}{I_n} P_{2\ell+1-n} P_n + \sum_{n=1}^{2\ell-1} \frac{n}{\alpha_n} X_{2\ell-n} X_n
  \right),
  \nn \\
  Q &= - \sum_{n=1}^{2\ell} \frac{n}{I_n} P_{2\ell-n} X_{n-1},
  \nn \\
  S &= - \sum_{n=1}^{2\ell} \frac{n}{I_n} P_{2\ell+1-n} X_{n-1}.
\end{align}
To verify this, note the following identities:
\begin{align}
   & \sum_{n=0}^{2\ell} \frac{1}{I_n} P_{2\ell-n} P_n = -\ell - \hf,
   \nn \\
   & \sum_{n=0}^{2\ell-1} \frac{1}{\alpha_n}  X_{2\ell-1-n} X_n = \ell.
\end{align}
This gives a star representation of $\tGl $ for both anti-involution \eqref{adjoint1} and \eqref{adjoint2}. However, this is not a superstar representation with \eqref{superadjoint1} since \eqref{superadjoint1} does not respect the relations \eqref{MassExt}. 

%
\subsection{Vector field realization of $\Gl$}

If a vector field realization of $\g_{\ell} $  is converted to the one for $\Gl$ , 
due to \eqref{PPdef} the realization of $ \Gl $ will be given by second order differential operators. It is, however, unusual to realize Lie superalgebras by second order operators. 
In this subsection we give a realization of $ \Gl$ in terms of first order differential operators of \Z2 generalization of the Grassmann numbers. 

A \Z2 generalization of Grassmann numbers was introduced in \cite{rw1,rw2}. The 
\Z2 graded numbers $\xi_{\bm{a}}, \xi_{\bm{b}} $ with degree $ \bm{a}, \bm{b} \in$ \Z2 are defined by the relation:
\beq
   \llbracket \xi_{\bm{a}}, \xi_{\bm{b}} \rrbracket =0. \label{Z2GrassDef}
\eeq
By exponential mapping one may introduce the color supergroup generated by $ \Gl$:
\begin{equation}
   G_{\ell} = \exp( \Gl )
\end{equation}
where the group parameters are given by the \Z2 graded numbers. 
It would be more helpful to use different notations for numbers with different degree. 
We use the following notation in the sequel:
\[
  (0,0) \ \; x,\; y, \; w \qquad (0,1) \ \; \psi \qquad (1,0) \ \; \theta,\; \sigma \qquad (1,1) \ \; z
\]
Note that $ \psi, \theta, \sigma $ are nilpotent and others are not. 
With these notations we parametrize the elements $g \in G_{\ell} $ as
\begin{align}
   g &= \exp(x_1H) \exp(\theta_1 Q) \exp\Big( \sum_{n=0}^{2\ell} \psi_n P_n \Big) 
      \exp\Big( \sum_{n=0}^{2\ell-1} z_n X_n \Big) 
      \exp(\theta_2 S) \exp(x_2 K) 
      \nonumber
   \\
    & \times \exp(x_3 D) \exp\Big(\sum_{n\leq m}^{2\ell} y_{nm} P_{nm} \Big) 
      \exp\Big( \sum_{n<m}^{2\ell-1} w_{nm} X_{nm} \Big)
      \exp\Big( \sum_{n=0}^{2\ell} \sum_{m=0}^{2\ell-1} \sigma_{nm} \Lambda_{nm} \Big),
      \label{parametrization}
\end{align}
where, reflecting the symmetries of the generators, 
$
   y_{nm} = y_{mn}, \ w_{nm} = -w_{mn}, 
$ 
and $ \sigma_{nm} $ is neither symmetric nor antisymmetric. 

 Now we consider the space of $ C^{\infty} $ class functions $ f(g) $ over $G_{\ell}.$ 
In order to define an action of $\Gl$ on this space of functions 
we need derivatives with respect to the \Z2 graded numbers.  
Such derivatives have been introduced as a natural generalization of that for Grassmann numbers \cite{NAJS}. 
Namely, the derivative $ \displaystyle \F{\xi_{\bm{a}}} $ is an operator acting on $f(g)$ from the left specified by the following properties:
\begin{enumerate}
   \renewcommand{\labelenumi}{(\roman{enumi})}
  \item $  \displaystyle \F{\xi_{\bm{a}}} $ has the degree same as $ \xi_{\bm{a}} $ and 
  $ \displaystyle \left\llbracket \F{\xi_{\bm{a}}}, \F{\eta_{\bm{a}}}  \right\rrbracket = 0$
  \item $  \displaystyle \F{\xi_{\bm{a}}} $ annihilates constant functions: 
  $  \displaystyle \F{\xi_{\bm{a}}} 1 = 0$
  \item it acts on degree one monomials in the same way as ordinary derivative:
   \begin{equation}
       \F{\psi_n} \psi_m = \delta_{nm}, \qquad
       \F{\sigma_{nm} } \sigma_{kr} = \delta_{nk} \delta_{mr}, \ \text{etc}
       \label{derivative1}
   \end{equation}
  \item when it acts on a monomial the variable is moved to the left most position, then \eqref{derivative1} is applied. For instance,
  \[
    \frac{\partial}{\partial\psi_2} x_2 \psi_1 \psi_2 = - x_2 \psi_1, 
   \quad 
  \frac{\partial}{\partial\theta_1} \psi_1 \theta_1 z_3 = \psi_1 z_3,
   \quad 
  \frac{\partial}{\partial z_1} x_2 \psi_3 z_1^2 = -2 x_2\psi_3 z_1.
  \]
\end{enumerate}   
   
  With this definition of derivatives, 
we define a left action of $\Gl$ on $ f(g) $ in the standard way of Lie theory: 
\beq
   Y f(g) = \left. \frac{d}{d\tau} f(e^{-\tau Y}g) \right|_{\tau=0}, \quad Y \in \Gl \label{LAdef}
\eeq
where $ \tau $ is a parameter having the degree same as $ Y.$ 
It is not difficult to verify that the left action gives a realization of $ \Gl $ on the space of functions $f(g). $ 
We give explicit formulae of the left action in the parametrization \eqref{parametrization}.
\begin{align}
   H &= -\F{x_1},
   \\
   D &= -x_1 \F{x_1} + x_2 \F{x_2} - x_3 \F{x_3} - \hf \theta_1 \F{\theta_1} 
      + \hf \theta_2 \F{\theta_2} 
      \nonumber
     \\
     & - \sum_{n=0}^{2\ell} (\ell-n) \psi_n \F{\psi_n} 
       - \sum_{n=0}^{2\ell-1} \Big( \ell - \hf -n\big) z_n \F{z_n},
\end{align}

\begin{align}
  K &= - 2x_1 D -x_1^2 \F{x_1} - (1 + \theta_1 \theta_2) \F{x_2} 
       - \theta_1 \F{\theta_2} 
    \nonumber
    \\
    & + \hf \sum_{n=0}^{2\ell} \sum_{m=0}^{2\ell-1}  
       \left\{
          \psi_n z_m \theta_1  \sfP^{n\;m+1}
          + (2\ell-1-m) \psi_n z_m \theta_1 \theta_2 \, \sfL^{n\;m+1}
       \right\}            
    \nonumber
   \\
   & + \sum_{n,m=0}^{2\ell-1} \left\{ -z_n z_m \sfP^{n+1\;m+1}
     -\hf  (2\ell-n) \psi_n z_m \theta_1\, \sfX^{nm}
     +  (2\ell-1-n) z_n z_m \theta_1 \theta_2 \,\sfX^{n+1\;m}
   \right\}
   \nonumber
   \\
  & + \sum_{n,m=0}^{2\ell-1} 
    (z_n z_m \theta_1 -\hf (2\ell-n) \psi_n z_m \theta_1 \theta_2 )\, \sfL^{n+1\;m},
\end{align}
where
\begin{align}
  \sfP^{rs} &= \sum_{i=0}^{2\ell-r} \sum_{j=0}^{2\ell-s} 
    \binom{2\ell-r}{i} \binom{2\ell-s}{j} (-x_2)^{i+j} 
    \exp\big( (r+s+i+j-2\ell) x_3 \big) \F{y_{r+i\;s+j}}, 
   \\
  \sfX^{rs} &= \sum_{i=0}^{2\ell-1-r} \sum_{j=0}^{2\ell-1-s} 
    \binom{2\ell-1-r}{i} \binom{2\ell-1-s}{j} (-x_2)^{i+j} 
    \exp\big( (r+s+i+j-2\ell+1) x_3 \big)
    \nonumber
    \\
    & \qquad \times \F{w_{r+i\;s+j}},
   \\
  \sfL^{rs} &= \sum_{i=0}^{2\ell-r} \sum_{j=0}^{2\ell-1-s} 
    \binom{2\ell-r}{i} \binom{2\ell-1-s}{j} (-x_2)^{i+j} 
    \exp\big( (r+s+i+j-2\ell+\hf) x_3 \big)
    \nonumber
    \\
    & \qquad \times \F{\sigma_{r+i\;s+j}},   
\end{align}

\begin{align}
  Q &= -\F{\theta_1} + \theta_1 \F{x_1},
  \\
  S &= -x_1 Q  + (2x_2 \theta_1 - \theta_2) \F{x_2} 
     - 2\theta_1 \F{x_3}   -(1+\theta_1 \theta_2) \F{\theta_2}
     - 2 \sum_{n=0}^{2\ell} (\ell-n) \theta_1 \psi_n  \F{\psi_n} 
      \nonumber
     \\
      &+ \sum_{n=0}^{2\ell-1}\Big\{\, z_n \F{\psi_{n+1}} 
          + \big( (2\ell-1-2n) z_n \theta_1 + (2\ell-n) \psi_n \big) \F{z_n} \,\Big\}
      \nonumber
    \\
    & +  \sum_{n=0}^{2\ell} \sum_{m=0}^{2\ell-1}  \left\{
      \hf \psi_n z_m  \sfP^{n\; m+1}
     + \big( (2\ell-n) \psi_n z_m \theta_2 + z_n z_m \big) \sfL^{m+1\;n}
     \right\}
     \nonumber
    \\
  & + \hf \sum_{n=0}^{2\ell} \sum_{m=0}^{2\ell-1} \left\{
      (2\ell-m-1) \psi_n z_m \theta_2 \, \sfL^{n\;m+1}  
    -  (2\ell-n) \psi_n z_m \theta_2 \, \sfL^{n+1\;m} 
  \right\}
  \nonumber
  \\
    & - \sum_{n,m=0}^{2\ell-1} \left\{
     z_n z_m \theta_2 \sfP^{n+1\;m+1} 
     + \hf (2\ell-m) \psi_m z_n \, \sfX^{nm}  
     + (2\ell-1-m) z_n z_m \, \sfX^{n\;m+1}
    \right\},
\end{align}

\begin{align}
  P_r &= \sum_{n=0}^r \BN{r}{n} \Big( -\F{\psi_n} + n \theta_1 \F{z_n} \Big)
      \nonumber
      \\
      & + \sum_{m=0}^r \BN{r}{m} \left\{
         \sum_{n=0}^{2\ell} 
            \left(  
               \Big(\hf + m \theta_1 \theta_2 \Big) \psi_n \, \sfP^{nm}
               + \hf (2\ell-m) \psi_n \theta_2 \,\sfL^{nm}
               - m \theta_1 \psi_n \sfL^{n\;m-1}
            \right)
         \right.
         \nonumber
         \\
      & 
       - \sum_{n=0}^{2\ell-1} m \Big( \hf \theta_1 z_n + (2\ell-n) \theta_1 \theta_2 \psi_n \Big) \sfX^{n\;m-1}
        \nonumber
        \\
       & \left.
         + \hf \sum_{n=0}^{2\ell-1} 
            \Big(
               \big( (2\ell-n) \psi_n \theta_2 + m \theta_1 \theta_2 z_n \big) \sfL^{mn}
              - m \theta_1 \theta_2 z_n \, \sfL^{n+1\; m-1}
            \Big)
      \right\},
\end{align}
where
\begin{equation}
  \BN{r}{n} = \binom{r}{n} (-x_1)^{r-n}.
\end{equation}

\begin{align}
   X_r &= 
      \sum_{n=0}^r \BN{r}{n} \left( -\F{z_n} + \theta_1 \F{\psi_n}  \right)
      \nonumber
      \\
      & + \sum_{m=0}^r \BN{r}{m} 
        \left\{
           \sum_{n=0}^{2\ell} \psi_n
             \left(
                -\hf \theta_1  \, \sfP^{nm} + \theta_2  \, \sfP^{n\;m+1}
                + \big(1-\hf (2\ell-m) \theta_1 \theta_2 \big) \, \sfL^{nm}
             \right)
       \right.
       \nonumber
       \\
       & 
        + \sum_{n=0}^{2\ell-1}
           \left(
             \Big( \hf z_n + (2\ell-n) \psi_n \theta_2 \Big) \sfX^{nm} 
             - \hf (2\ell-n) \theta_1 \theta_2 \psi_n \sfL^{mn}
           \right)
        \nonumber
        \\
       & \left.
       - \hf \sum_{n=0}^{2\ell-1} z_n \theta_2 \big( \sfL^{n+1\;m} - \sfL^{m+1\;n} \big)
        \right\},
\end{align}

\begin{align}
  P_{rs} &= \sum_{n=0}^r\sum_{m=0}^s \Gamma^{rs}_{nm} 
    \left\{
       -(1+(n+m) \theta_1 \theta_2) \sfP^{nm}
    \right.
    \nonumber
    \\
    & 
     + \theta_1 \theta_2 \big( n(2\ell-m)\, \sfX^{m\;n-1} + m(2\ell-n)\, \sfX^{n\;m-1} \big)
     \nonumber
     \\
    & \left.
      + \theta_1 \big( n \, \sfL^{m\;n-1} + m \, \sfL^{n\;m-1} \big)
      - \theta_2 \big( (2\ell-m) \sfL^{nm} + (2\ell-n) \sfL^{mn}  \big) 
    \right\},
\end{align}
where
\begin{equation}
 \Gamma^{rs}_{nm} = \binom{r}{n}\binom{s}{m} (-x_1)^{r+s-n-m}.
\end{equation}

\begin{align}
  X_{rs} &= \sum_{n=0}^r\sum_{m=0}^s \Gamma^{rs}_{nm} 
    \left\{
      \theta_1 \theta_2 \big( \sfP^{m\;n+1} - \sfP^{n\;m+1} \big)
      - \big( 1 + (n+m-4\ell) \theta_1 \theta_2 \big) \sfX^{nm}
     \right.
     \nonumber
     \\
   & \left.
      + \theta_1 \big( \sfL^{nm}- \sfL^{mn}  \big)
      + \theta_2 \big( \sfL^{n+1\;m} - \sfL^{m+1;n} \big)
    \right\},
\end{align}

\begin{align}
  \Lambda_{rs} &= \sum_{n=0}^r\sum_{m=0}^s \Gamma^{rs}_{nm} 
    \left\{
      \theta_1 \sfP^{nm}  + \theta_2 \sfP^{n\;m+1}
      + n \theta_1 \, \sfX^{n-1\;m}  - (2\ell-n) \theta_2 \, \sfX^{nm}
    \right.
    \nonumber
    \\
   & \left.
   -\big( 1 + (n+m-2\ell) \theta_1 \theta_2\big) \sfL^{nm}
   + (2\ell-n) \theta_1\theta_2 \, \sfL^{mn}
   + n \theta_1\theta_2 \, \sfL^{m+1\;n-1}
    \right\}.
\end{align}

One may repeat the same computation for $\tGl $ and obtain its vector field representation. 
However, the computation is so lengthy and cumbersome. We thus decided not include it in this paper.

%
%
%
\setcounter{equation}{0}
\section{Concluding remarks}
\label{SEC:CR}

We have introduced two classes of novel color superalgebras of \Z2 grading and studied their
properties and, to some extent, their representations. 
Our strategy of construction of a color superalgebra is to realize it in the enveloping algebra of
${\cal N}=1$ SCGA. 
This is a suitable approach because it is also applicable to higher $ \cal N $ SCGA and other Lie
(super)algebras. 
For instance, in \cite{NAJS} a color superalgebra is constructed by realizing it in the enveloping
algebra of ${\cal N}=2$ super Schr\"odinger algebras and in \cite{NA2}, a Heisenberg algebra of
multi-mode boson operators is used to construct a color superalgebra.  Furthermore, one may apply
this procedure to construction of infinite dimensional color superalgebras \cite{NaPsiJs}.  Of
course, not all color superalgebras are realized in the enveloping algebra of some Lie
(super)algebra.  Such an example is found in \cite{NAJS}. 

We give a vector field representation of $\Gl.$ This is an important step in order to consider physical
applications of $\Gl$ and geometric aspects of color  supergroups.  Regarding representation theory,
representations of the highest weight type are also important in many aspects. The algebras $\Gl$
and $\tGl$ admit a triangular-type vector space decomposition. Other color superalgebras discussed
in the literature also admit such a decomposition. However, it seems that we need some care to discuss Verma
modules over the color superalgebras \cite{NA1}. This will be important future work. 

One of the motivations of the present work is to find physical systems which have some relation to
a color superalgebra. 
One way of achieving this is to start with physical systems having (dynamical) symmetry generated by SCGA.
Then we try to extend the symmetry to a color superalgebra 
as in the case of the L\'evy-Leblond equation \cite{aktt1,aktt2}.

%
%
%
\section*{Acknowledgements}

N.A. and J.S. would like to thank P.S. Isaac for his warm hospitality at The University of Queensland, where
most of this work was undertaken. J.S. would also like to acknowledge the support of the School of Mathematics and
Physics at The University of Queensland for the Ethel Raybould Fellowship.

%
%
%

\end{document}